 \documentclass[preprint2]{aastex}

\usepackage{epsfig}
\usepackage{epstopdf}
\usepackage{graphicx}

\slugcomment{Submitted to ApJL on 2013, May, 8th ; Accepted on 2013, May, 24th}

\shorttitle{A probable giant planet around HD\,95086}
\shortauthors{Rameau et al.}

\begin{document}


\title{Discovery of a probable $4-5$ Jupiter-mass exoplanet to HD\,95086 by direct-imaging}


\author{J. Rameau\altaffilmark{1}, G. Chauvin\altaffilmark{1}, A.-M. Lagrange\altaffilmark{1}, A. Boccaletti\altaffilmark{2}, S. P. Quanz\altaffilmark{3}, M. Bonnefoy\altaffilmark{4}, J. H. Girard \altaffilmark{5}, P. Delorme\altaffilmark{1}, S. Desidera\altaffilmark{6},  H. Klahr\altaffilmark{4}, C. Mordasini \altaffilmark{4}, C. Dumas\altaffilmark{5}, M. Bonavita\altaffilmark{6}, T. Meshkat\altaffilmark{7}, V. Bailey\altaffilmark{8}, and M. Kenworthy\altaffilmark{7}}


\altaffiltext{}{Submitted to ApJL on 2013, May, 8th ; Accepted on 2013, May, 24th\\
Electronic adress: julien.rameau@obs.ujf-grenoble.fr\\
Based on observations collected at the European Organisation for Astronomical Research in the Southern Hemisphere, Chile, under programs number 087.C-0292, 088.C.0085, 090.C-0538, 090.C-0698, and 090.C-0728.}
\altaffiltext{1}{UJF-Grenoble 1 / CNRS-INSU, Institut de Plan\'etologie et d'Astrophysique de Grenoble (IPAG) UMR 5274, Grenoble, F-38041, France}
\altaffiltext{2}{LESIA, Observatoire de Paris, CNRS, University Pierre et Marie Curie Paris 6 and University Denis Diderot Paris 7, 5 place Jules Janssen, 92195 Meudon, France}
\altaffiltext{3}{Institute for Astronomy, ETH Zurich, Wolfgang-Pauli-Strasse 27, 8093 Zurich, Switzerland}
\altaffiltext{4}{Max Planck Instiute f\"ur Astronomy, K\"onigsthul 17, D-69117 Heidelberg, Germany}
\altaffiltext{5}{European Southern Observatory, Casilla 19001, Santiago 19, Chile}
\altaffiltext{6}{INAF - Osservatorio Astronomico di Padova, Vicolo dell' Osservatorio 5, 35122, Padova, Italy}
\altaffiltext{7}{Leiden Observatory, Leiden University, P.O. Box 9513, 2300 RA Leiden, the Netherlands}
\altaffiltext{8}{Steward Observatory, Department of Astronomy, University of Arizona, 933 North Cherry Avenue, Tucson, AZ 85721-0065, USA}


\begin{abstract}
Direct imaging has just started the inventory of the population of gas giant planets on wide-orbits around young stars in the solar neighborhood. Following this approach, we carried out a deep imaging survey in the near-infrared using VLT/NaCo to search for substellar companions. We report here the discovery in L\,' ($3.8~\mu m$) images of a probable companion orbiting at $56~$AU the young ($10-17~$Myr), dusty, and early-type (A8) star HD 95086. This discovery is based on observations with more than a year-time-lapse. Our first epoch clearly revealed the source at $\simeq10~\sigma$ while our second epoch lacked good observing conditions hence yielding a $\simeq3~\sigma$ detection. Various tests were thus made to rule out possible artifacts. This recovery is consistent with the signal at the first epoch but requires cleaner confirmation. Nevertheless, our astrometric precision suggests the companion to be comoving with the star, with a $3~\sigma$ confidence level. The planetary nature of the source is reinforced by a non-detection in Ks-band ($2.18~\mu m$) images according to its possible extremely red Ks - L\,' color. Conversely, background contamination is rejected with good confidence level. The luminosity yields a predicted mass of about $4-5~\mathrm{M}_\mathrm{Jup}$ (at $10-17~$Myr) using "hot-start" evolutionary models, making HD\,95086\,b the exoplanet with the lowest mass ever imaged around a star.
\end{abstract}





\section{Introduction \label{sec:intro}}
Direct imaging is challenging to search for orbiting giant planets due to the high planet-star contrast and small angular separation explored. As a result, very few
planetary-mass companions have been detected by direct imaging, initially at
relatively large separations ($\ge 100~$AU) around solar-type to low-mass stars (e.g., \citealt{chauvin05}, \citealt{bejar08}, \citealt{lafreniere08}, and \citealt{ireland11}). They may have
a relatively high mass-ratio ($q\sim0.2-0.02$) suggesting a stellar-like
formation origin. More recently, the discoveries of giant planets around the young and dusty early-type stars HR\,8799 \citep{marois08,marois10} and
$\beta$ Pictoris \citep{lagrange10} at smaller physical separations ($\le 70$~AU) and with lower mass-ratio ($q\sim0.002$) suggested rather a formation within the circumstellar disk
either by core accretion \citep{pollack96} or gravitational instability \citep{cameron78}. Fomalhaut b \citep{kalas08} is a peculiar case since its photometry seems to be contaminated by reflected light from the dust \citep{currie12}, making the precise determination of its mass more uncertain. Also, the (proto-) planet candidates around LkCa\,15 \citep{kraus12} and HD\,100546 \citep{quanz13} still require confirmations.\\
Consequently, every single discovery has a tremendous impact
on the understanding of the formation, the dynamical evolution, and
the physics of giant planets.

Although very few directly imaged giant planets, still very massive, have been reported in the literature, so far only one (maybe HR 8799\,b, e.g., \citealt{marois10}), with mass lower than $7~\mathrm{M}_\mathrm{Jup}$, has been imaged around a star. We report here the discovery of a probable $4-5~\mathrm{M}_\mathrm{Jup}$ giant planet around HD\,95086, the exoplanet with the lowest mass ever imaged around a star. If the comoving status of the companion is confirmed, this giant planet may become a benchmark for physical studies of young giant planets but also for formation and evolution theories of planetary systems.





\section{Observations and data reduction\label{sec:obs}}

\subsection{The star HD\,95086}
HD\,95086 was identified as an early-type member of the Lower Centaurus Crux (LCC) association by \citet{zeeuw99} and also by \citet{madsen02}. The membership was established on the grounds that the star shares a similar velocity vector in the galactic framework with other LCC members. HD\,95086 has a distance of $90.4\pm3.4~$pc \citep{leeuwen07}, which is approximately the mean distance of the LCC association.
About the age of the association, \citet{mamajek02} followed by \citet{pecaut12} derived $17\pm2~$Myr (based on isochrone fitting) while \citet{song12} derived $\simeq 10~$Myr (by comparison with Lithium equivalent width of members of other nearby associations). Systematic differences between these two fully independent methods may be responsible for the discrepancy. Nonetheless, the full assessment of the age is beyond the scope of this paper and adopting $10$ or $17~$Myr has only a limited impact in the following results.
\newline
\citet{houk75} proposed HD\,95086 to have a class III luminosity and a A8 spectral-type with a mass of $\simeq1.6~\mathrm{M}_\odot$. However, its good-quality trigonometric parallax and thus derived luminosity and effective temperature undoubtedly place it close to zero-age main sequence stars of LCC and thus reject the super-giant phase.
\newline
Another interesting property of HD\,95086 is that observations in the mm \citep{nilsson10} and in the mid-to-far infrared (\citealt{rizzuto12}, \citealt{chen12}) revealed
a large dust-to-star luminosity ratio
($\mathrm{L}_\mathrm{d}/\mathrm{L}_\star=10^{-3}$), indicating the presence of a so far unresolved debris disk.
\newline
Finally, \citet{kouwenhoven05} observed HD\,95086 with adaptive-optics and identified a background star with a separation of $4.87\,''$ and position angle of $316~$deg to the star (based on its K magnitude).

\subsection{Observations}

\begin{table*}[t!]
\caption{Observing log of HD\,95086 with VLT/NaCo. \label{tab:log}}
\centering
\footnotesize
\begin{tabular}{llllllllll}     
\tableline
\tableline
Type & Date & Cam./Filter & DIT $\times$ NDIT & N$_{\rm exp}$ & $\pi$-start/end& $\langle$Airmass$\rangle$\tablenotemark{a} & $\langle$FWHM$\rangle$\tablenotemark{a} & $\langle\tau_0\rangle$\tablenotemark{a} & $\langle\mathrm{E}_\mathrm{c}\rangle$\tablenotemark{a} \\
        & &  & (s)  & &(deg) &  & ($\,\!''$) & (ms) & ($\%$)\\
\tableline
$\theta_1$ Ori C & 2011/12/18 & L27/L\,' & 0.2 $\times$ 150 & 6 & --/-- & 1.11 & 0.78 & 7.4  & 45.9 \\
PSF & 2012/01/11& L27/L\,'+ND & 0.2 $\times$ 80 & 10 & -9.32/-8.19 & 1.39 & 0.75 & 3.6& 61.1 \\
Deep & 2012/01/11 & L27/L\,' & 0.2 $\times$ 100 & 156 & -7.59/16.96 & 1.39 & 0.76 & 3.5& 58.2\\
$\theta_1$ Ori C & 2012/02/10 & L27/L\,' & 0.2 $\times$ 195 & 10 & --/-- & 1.06 & 0.50 & 6.0  & 37.1 \\
\tableline
PSF & 2013/02/14 & S13/Ks  & 0.2 $\times$ 100 & 4 &  -45.18/-44.72  & 1.52 & 1.08 & 1.1&  54.5\\
Deep & 2013/02/14 & S13/Ks  & 0.5 $\times$ 100 & 88 &  -44.14/-14.68  & 1.45 & 1.08 & 1.1 & 22.4\\
\tableline
$\theta_1$ Ori C & 2013/03/24 & L27/L\,' & 0.2 $\times$ 50 & 10 & --/--  & 1.16 & 1.56 & 5.9  & 52.1 \\
Deep & 2013/03/14 & L27/L\,'  & 0.2 $\times$ 100 & 162 &  3.20/28.18  & 1.41 & 1.77 & 1.0& 37.2\\
PSF & 2013/03/14 & L27/L\,'+ND  & 0.2 $\times$80 & 10 &  29.61/30.68  & 1.44 & 1.65 & 0.9 & 32.1\\
\end{tabular}
\tablecomments{''ND'' refers to the NaCo ND\_Long filter (transmission of $\simeq 1.79~\%$), "PSF" (point-spread function) for unsaturated exposures, ''Deep'' to deep science observations, "DIT" to exposure time, and $\pi$ to the parallactic angle at start and end of observations. $\theta_1$ Ori C was observed in field-stabilized mode.}
\tablenotetext{a}{The airmass, the FWHM, the coherence time $\tau_0 $ and energy $\mathrm{E}_\mathrm{c}$ are estimated in real time by the adaptive-optics system.}
\end{table*}

HD\,95086 was observed with VLT/NaCo ({\citealt{lenzen03}, \citealt{rousset03}) in thermal infrared with angular differential imaging \citep[ADI][]{marois06} mode as part of our direct-imaging survey of young, dusty, and early-type stars \citep{rameau13}. NaCo was configured for the L\,' filter ($\lambda_0 = 3.8~\mu m$, $\Delta \lambda=0.62~\mu m$) with the L27 camera in service mode for observations in 2012, January. The source was dithered within the instrument field-of-view ($\simeq14\,''\times14\,''$) in order to properly estimate and remove the background contribution. An observing sequence was made up of a first short set of unsaturated exposures to serve as calibrations for the point-spread function and for the relative photometry and astrometry. The sequence was followed by a one hour set of deep science-observations.

In 2013, follow-up observations were done and included two runs: one in February (back-up star, visitor mode, very bad conditions), with the Ks filter ($\lambda_0 = 2.18~\mu m$, $\Delta \lambda=0.35~\mu m$) and the S13 camera (platescale $\simeq 13.25$~mas/pixel), and one in March at L\,' with the L27 camera in service mode. Table \ref{tab:log} summarizes the observing log for each run. We recall HD\,95086 has $\mathrm{V}=7.36\pm0.01~$mag, $\mathrm{Ks}=6.79\pm0.02~$mag, and $6.70\pm0.04~$mag at $3.4~\mu m$ from the WISE database.

Finally, an astrometric calibrator, $\theta_1$ Ori C field, was observed for each observing run.

\subsection{Data reduction}

Data reduction (flat-fielding, bad pixel and sky removal, registration\footnote{The central star was not saturated in the deep science-observations; this thus enabled to get a very good accuracy on the measurements of the star position using Moffat fittings.}, and image selection) was performed using the IPAG-ADI pipeline (e.g., \citealt{lagrange10}, \citealt{chauvin12}, \citealt{rameau13}, and refs. therein). Stellar-halo subtraction was also done using all the ADI algorithms implemented in the pipeline: cADI, sADI \citep{marois06}, and LOCI \citep{lafreniere07}. The frames were finally de-rotated and mean-combined. The astrometry and photometry of any detected point source as well as their error estimates were done similarly as in \citet{chauvin12} and \citet{lagrange12b} by injecting fake planets using the unsaturated point-spread function reduced images. The noise per pixel was derived from the standard deviation calculated in a ring of $1.5$ FWHM width, centered on the star, with a radius of the separation of the source, and masking the point-source itself. The flux of the point-source was integrated over an aperture of $1.5$ FWHM in diameter. The final signal-to-noise (S/N) was then calculated on the same aperture size considering the noise per pixel and the aperture size in pixels.

The $\theta_1$ Ori C field data was reduced for detector calibrations by comparison with \textit{HST} observations by \citet{mccaughrean94} (using the same set of stars TCC0051, 034, 029, and 026 at each epoch). We found, for the two 2012 and 2013 data, a true north of $-0.37\pm0.02~$deg, $-0.38\pm0.03~$deg, and $-0.45\pm0.09~$deg  respectively; and a platescale of $27.11\pm0.06~$mas, $27.10\pm0.03~$mas,  and $27.10\pm0.04~$mas respectively.

Finally, the detection performance was derived by measuring the $5\sigma~$level noise in a sliding box of $5\times5~$pixel toward the direction of the point source and corrected for flux loss. The contrast was converted to mass with the "hot-start" COND models of \citet{baraffe03}.

Two independent pipelines (\citealt{boccaletti12}, \citealt{amara12}) were also used for consistency and error estimates.






\section{A companion candidate and a background star}
\subsection{Astrometry}


\begin{figure*}[th]
\epsscale{1.6}
\plotone{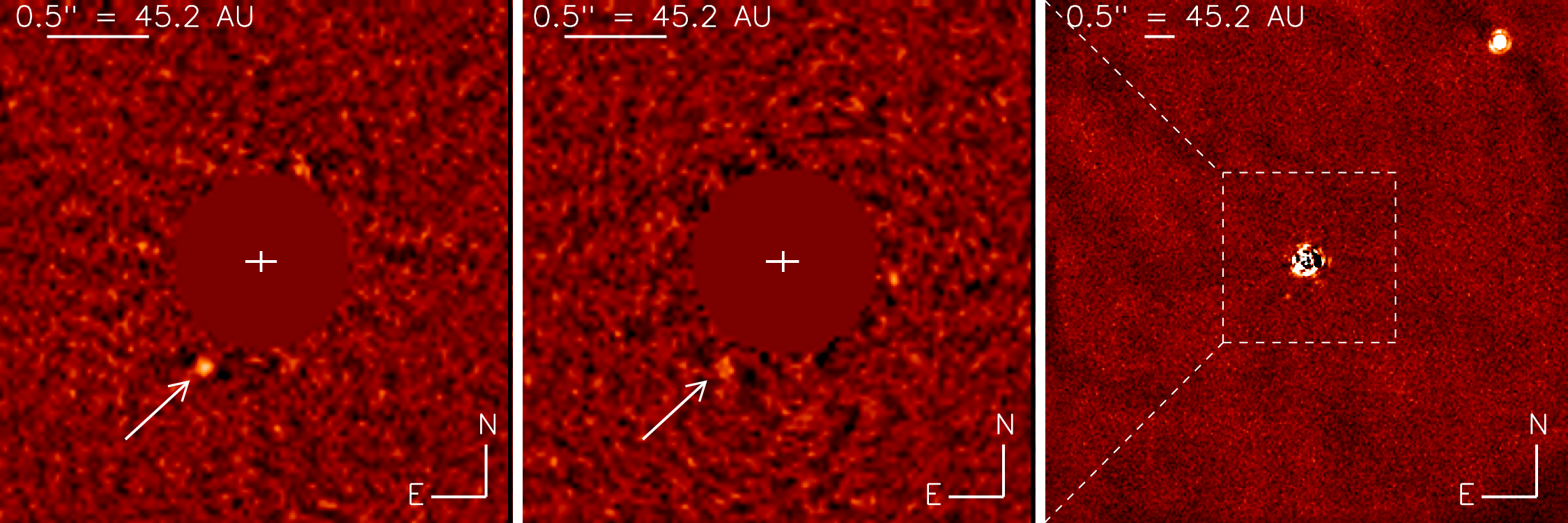}\\
\plotone{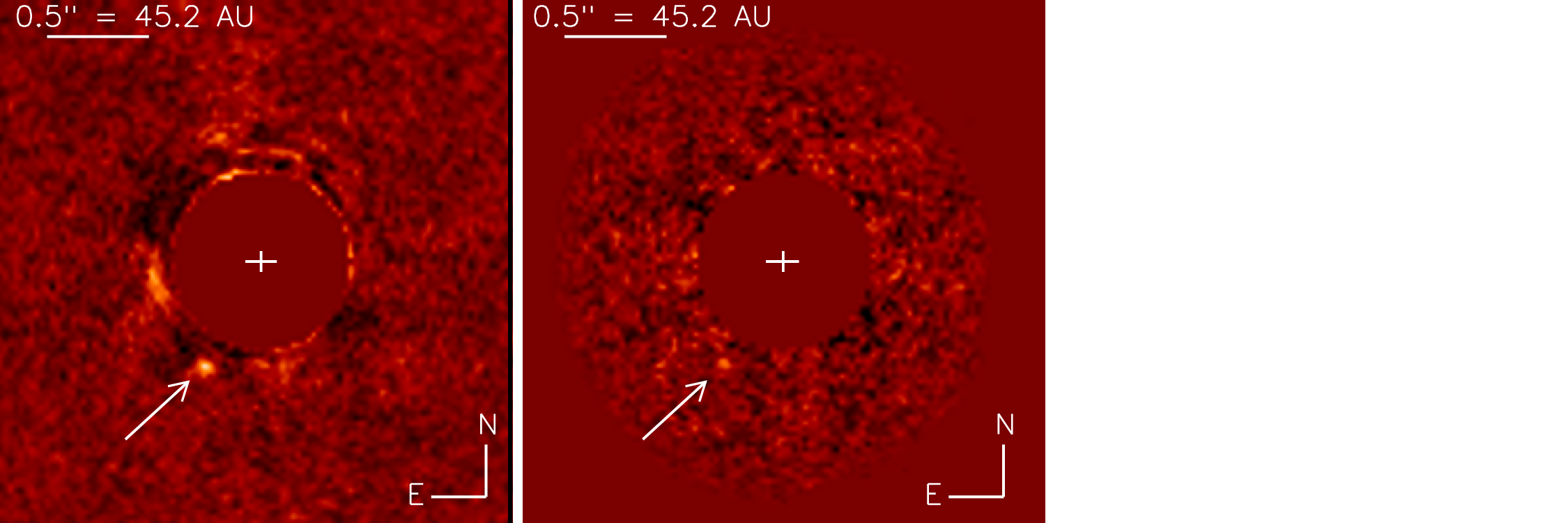}
\caption{Residual maps (sADI at top, LOCI at bottom) at L\,' showing the companion candidate in 2012 (\textbf{left}) with a S/N of nine and in 2013 (\textbf{middle}) with a S/N of three. \textbf{Top-right:} Larger field-of-view with the visual binary at northwest in 2012. Bright residuals are remaining speckles from the Airy ring at the same separation of the companion candidate.\label{fig:images}}
\end{figure*}

The data in 2012, January showed the background star detected by \citet{kouwenhoven05} at a projected separation of $4.540\pm0.015~\!''$ from the central star and a position angle of $319.03\pm0.25~$deg (see Fig. \ref{fig:images}, right panel). We also detected an additional fainter signal southeast of the star with a S/N of nine. The robustness of this detection was strengthened thanks to its systematic confirmation via the mean of a series of tests using: three independent pipelines, all our published flavors of ADI algorithms, and an extensive parameter space exploration. The source successfully passed all tests. Fig. \ref{fig:images}, \textbf{top- and bottom-}left panel, presents the companion candidate (CC) located at a separation of $623.9\pm7.4~$mas and position angle of $151.8\pm0.8~$deg from the central star, using the sADI algorithm (with 20 frames combined for $N_\delta=1$ (FWHM) at $r=540\,''$) and LOCI (with $N_\delta=0.75$ (FWHM), $dr=3$ (FWHM), $g=1$, and $N_A=300$ (FWHM)), respectively.
\newline
HD\,95086 being at very low galactic latitude ($b\simeq-8~$deg), the contamination by background objects is relatively high, even at L\,'. A time-lapse long enough with another dataset was mandatory to prove the companionship of each object (see Section \ref{sec:bkg}).

In the 2013 L\,' data, both objects were detected at similar locations as in 2012. However, the weather conditions strongly varied over the sequence thereby revealing the CC with a lower S/N of three. We carried out the same tests as for the 2012 data for this dataset and the CC was detected. The signal is thus consistent with the object seen in 2012, even at low S/N. The residual map with the CC southeast from the star is displayed in Fig. \ref{fig:images}, top- and bottom-central panel using sADI with the 2012 parameters and LOCI with $N_\delta=0.75$ (FWHM), $dr=1$ (FWHM), $g=0.5$, and $N_A=600$ (FWHM). The positions are $626.11\pm12.8~$mas and $150.7\pm1.3~$deg for the separation and position angle of the CC, and $4.505\pm0.016~\!''$ and $319.42\pm0.26~$deg for the background star.

At Ks, we did not detect the CC (Fig. \ref{fig:pm}, bottom-left panel). Conversely, the background star was revealed as well as seven other point sources not seen at L\,'. This may be consistent with background objects.

\subsection{Photometry}
 
In 2012 L\,' data, we derived the star-to-CC contrast to be $9.79\pm 0.40~$mag ($\mathrm{L\,'}=16.49\pm0.50~$mag) and $6.2\pm 0.2~$mag for the background star. The error budget includes, from high-to-low significance: photometry of the CC, neutral density, PSF flux estimate and variability.
Similar results have been obtained with other algorithms and pipelines. In 2013 L\,' data, the weakness of the CC-signal impacted to the estimation of the photometry and higher uncertainties than in 2012 data but the L\,' contrast was consistent and equals to $9.71\pm0.56~$mag. For the background star, we found $\Delta \mathrm{L\,'}=6.1\pm 0.2~$mag.

Finally, at Ks, we estimated $\Delta \mathrm{Ks}=5.84\pm 0.1~$mag for the background star. The non-detection of the CC directly also provided an lower-limit to the Ks - L\,' color of $1.2~$mag.

\section{Discussion}

In the following, we discuss the nature of the two detected objects to HD\,95086 as well as their physical properties.

\subsection{Background objects? \label{sec:bkg}}

\begin{figure}[h!]
\epsscale{1.1}
\plotone{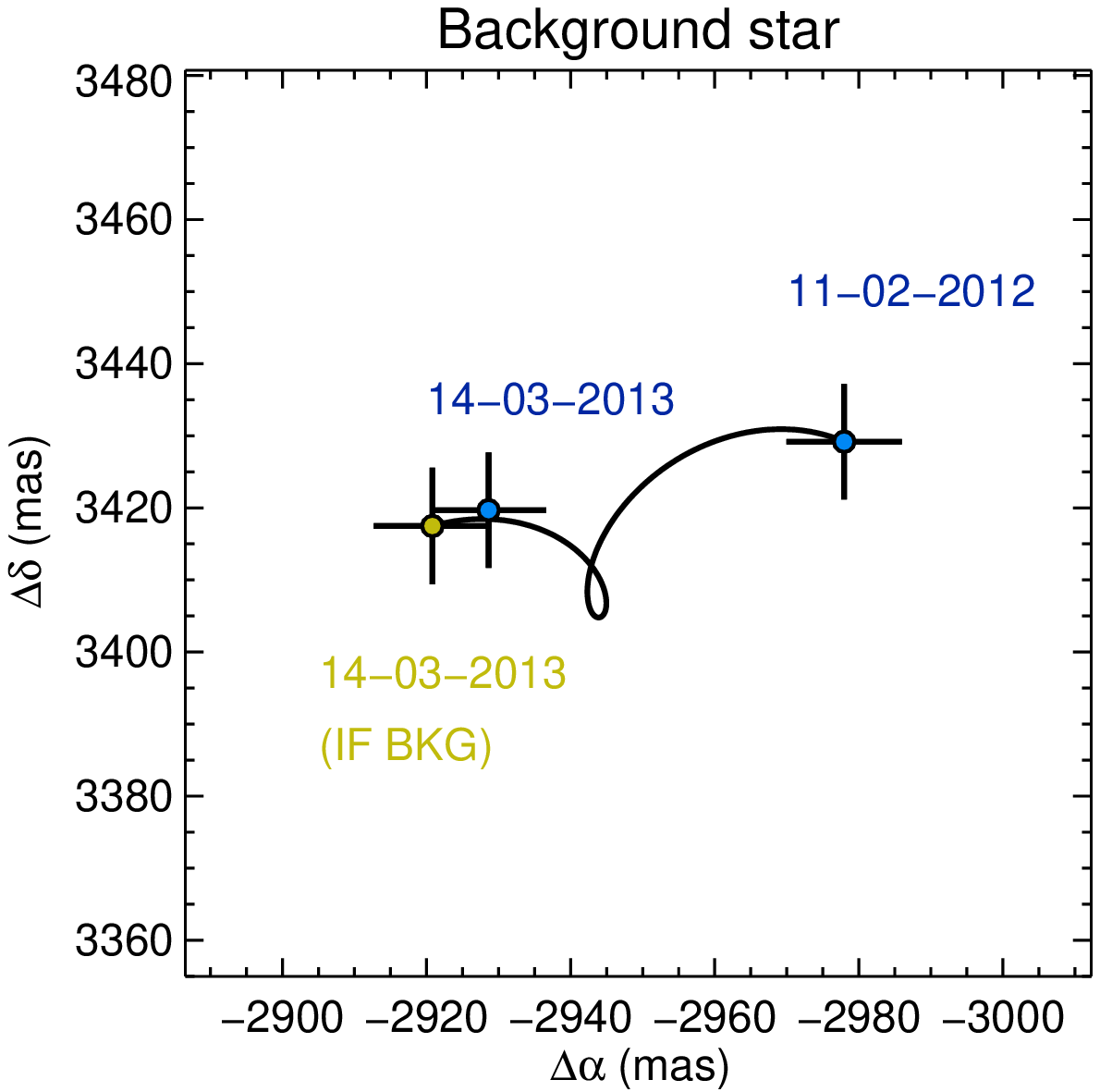}\\
\plotone{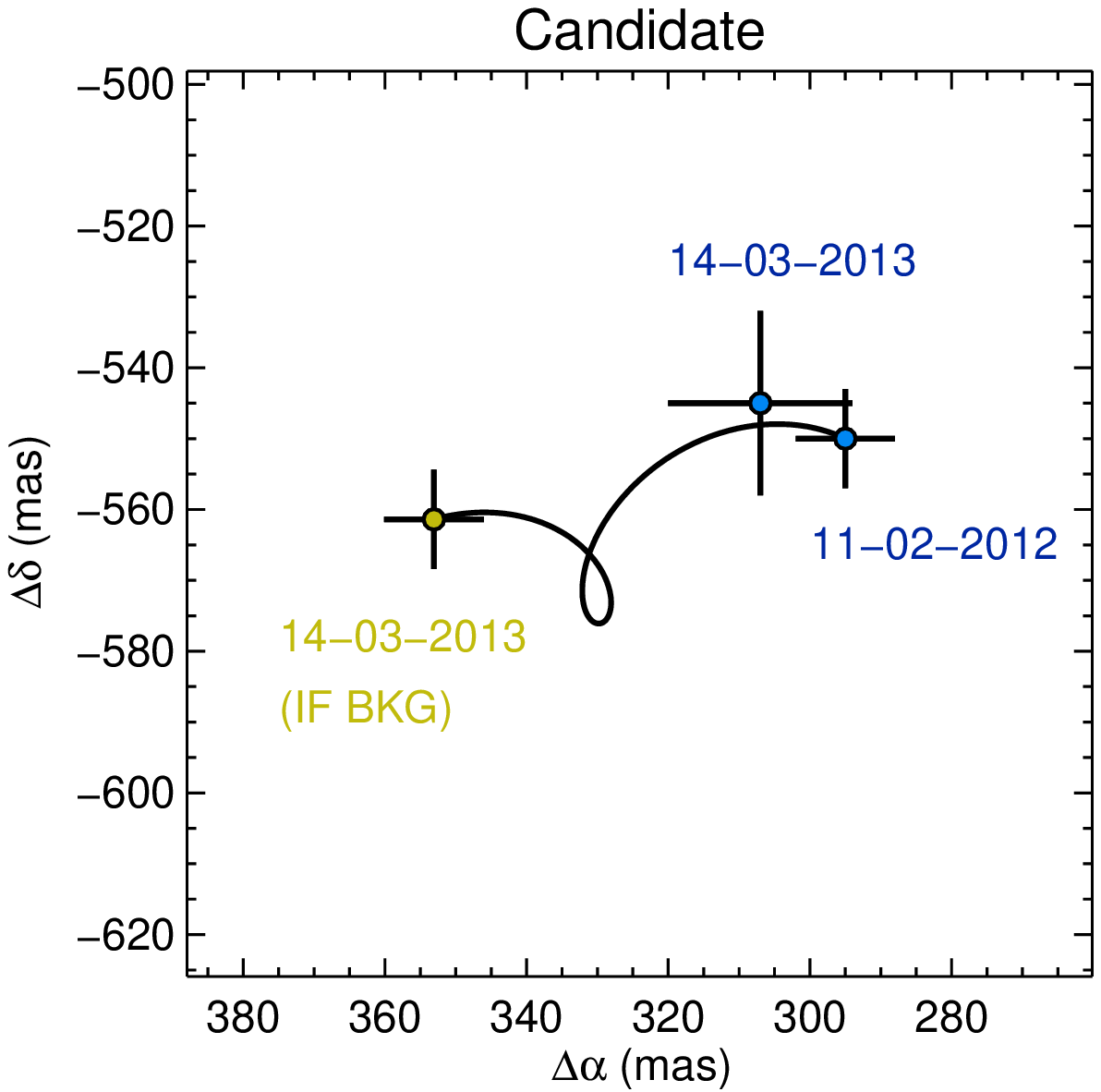}
\caption{Relative separations between the central star and a candidate companion, in right ascension ($\alpha$) and declination ($\delta$). The epoch-one astrometric point is plotted in blue (2012/01/11) and linked to the expected position of the CC, if it was a background object (gold, 2013/03/14), by a proper and parallactic motion track. The epoch-two astrometric point in 2013 is over plotted in blue. \textbf{Bottom:} Case of the CC which may be inconsistent with a background status (at the $3\sigma$ confidence level). \textbf{Top:} Case of the northwestern background star.\label{fig:pm}}
\end{figure}

\begin{figure*}[t!]
\epsscale{1.2}
\plotone{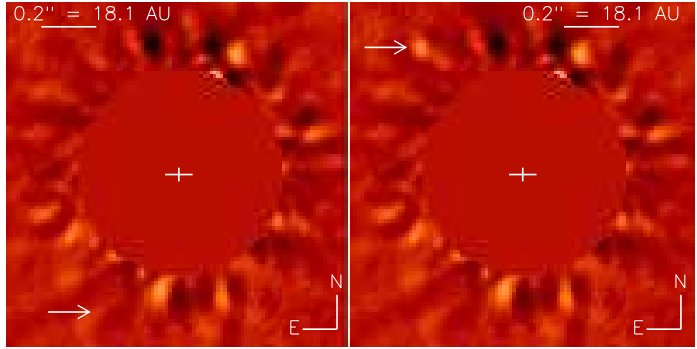}
\caption{Residual maps (sADI) at Ks showing the non-detection of the CC in 2013 (\textbf{left}) and the recovery of a M-dwarf contaminant with $\mathrm{Ks}-\mathrm{L\,'}\simeq0.4$ if located at the same separation as the CC (\textbf{right}). Speckles and residuals inner at and at the same separation as the CC are due to spiders and poor PSF subtraction. \label{fig:KsM}}
\end{figure*}

HD\,95086 has a proper motion of [$-41.41\pm0.42$, $12.47\pm0.36$] mas/yr and a parallax of $11.06\pm0.41$~mas \citep{leeuwen07} hence translating into an amplitude of $58.886\pm 0.002~$mas (more than 2 NaCo/L27 pixels) between the two epochs. Fig. \ref{fig:pm}, top-left panel, shows that the northwestern star is unambiguously of background, based on astrometric measurements rather than on photometric ones \citep{kouwenhoven05}. This analysis strengthened our capability to discriminate between background behavior or common proper motion with the parent star, despite the relatively low amplitude between the two epochs.

Fig. \ref{fig:pm}, top-right panel, presents the sky relative-positions for the CC. Its background nature may be excluded with a $\chi^2$ probability of $3.10^{-3}$ ($3\sigma$ confidence level). For farther investigations to the background hypothesis, we ran simulations with the Besan\c con galactic model \citep{robin03} to identify the probability of contamination by stars with $\mathrm{L\,'}\simeq17~$mag and their properties. We found that in a field-of-view of radius of $1\,''$ around HD\,95086, this probability is about $0.11~\%$ and dominated by M dwarfs (peak at $\mathrm{Ks}=19~$mag). Making the assumption of a M-type background star, the resulting $\mathrm{Ks}-\mathrm{L\,'}\simeq0.4$ color would imply $\mathrm{Ks}\simeq16.9~$mag, which would easily be detected in our observations. Therefore, we injected a point-spread function scaled to the flux into the Ks data at the separation of the CC and reduced within the pipeline. The signal was unveiled with a S/N of fifteen (see Fig. \ref{fig:KsM}, right panel). As a result, the Ks dataset should have exhibited the CC if it had been the reddest contaminant even though the observing conditions were bad. For these reasons, background contamination by a late-K to M dwarf seems improbable and only a very red object (planet, brown-dwarf) may match the Ks - L\,' constraint. Finally, according to \citet{delorme10}, the probability of finding a fore/background field L or T dwarf around HD\,95086 ($1\,'$' radius) down to L\,'$\le 17~$mag is about $10^{-5}$. 

From both astrometry and photometry, we conclude that a contamination appears very unlikely.

\subsection{Physical properties of the candidate companion and additional planets \label{sec:prop}}

In the following, we assume the CC is a bound companion named hereafter HD\,95086\,b.

Firstly, the measured separation of $623.9\pm 7.4~$mas of HD\,95086\,b (from the 2012 data with the highest S/N) translates into a projected distance of $56.4\pm 0.7$ AU.

\begin{figure}[h!]
\epsscale{1}
\plotone{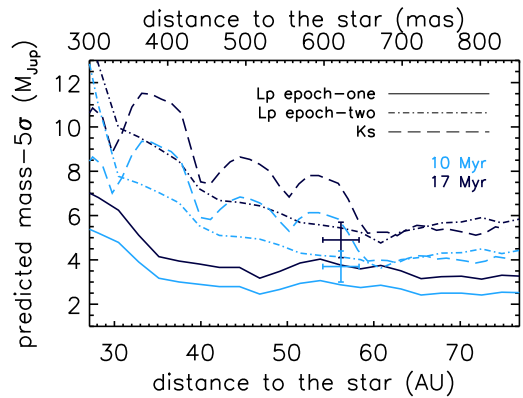}
\caption{$5\sigma$ detection limits, towards the direction of the CC, in mass vs separation for each epoch in L$\!'$ (solid, 2012, and dash-dotted, 2013, lines) and Ks (dashed line). Limits have been derived for an age of $17~$Myr (light color) and $10~$Myr (dark color) from COND models. The CC properties have been over plotted at each age. Note that for the second epoch in L\,', it is detectable at a S/N of three only, thus below the limit here. We also see from the Ks performance that the CC is indeed not detectable.\label{fig:limmass}}
\end{figure}

\begin{figure}[h!]
\epsscale{1}
\plotone{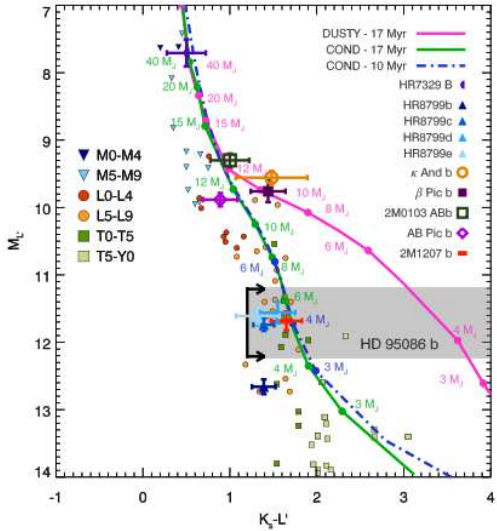}
\caption{$\mathrm{M}_\mathrm{L\,'}$ versus Ks - L\,' magnitude-color diagram. The location of HD\,95086\,b with the lower-limit at Ks is shown with the shaded region with respect to field M (triangles), L (circles), and T (squares) dwarfs \citep{leggett13}. Colors of young substellar companions are overlaid as well as COND and DUSTY evolutionary tracks at $10$ (dashed-line) and $17~$Myr (solid-lines). \label{fig:col-mag}}
\end{figure}

Given the observed contrast ($\Delta\mathrm{L\,'}=9.79\pm 0.40~$mag), distance ($90.4\pm3.4~$pc), and HD\,95086 magnitude ($6.70\pm 0.09$), we derived $\mathrm{M}_{\mathrm{L\,'}}=11.71\pm 0.53~$mag for HD\,95086\,b. This translates, according to the COND evolutionary model \citep{baraffe03}, into a mass of $5\pm 1~\mathrm{M}_{\mathrm{Jup}}$ at $17\pm2~$Myr, and $4\pm 1~\mathrm{M}_{\mathrm{Jup}}$ at $10~$Myr. We checked that given such a mass, HD\,95086\,b cannot be detected at Ks. Fig. \ref{fig:limmass} displays L\,' $5\sigma$-detection performances toward the direction of HD\,95086\,b. Our sensitivity ruled out any additional companion as light as $4~\mathrm{M}_\mathrm{Jup}$ from $48~$AU and $1200~$AU. The presence of any planet more massive than $8~\mathrm{M}_{\mathrm{Jup}}$ and $12~\mathrm{M}_{\mathrm{Jup}}$ can be excluded beyond $38~$AU and $34~$AU respectively, in projected separation. Moreover, we attempted a comparison of the L' band magnitude of HD95086 b to "warm-start" evolutionary models predictions \citep{SB12} with different initial conditions (see Fig. 11 of \citealt{bonnefoy13}). We found a mass greater than  $3~\mathrm{M}_{\mathrm{Jup}}$ using the youngest age estimate of the system and three-times solar metallicity hybrid cloud models. The lack of prediction for $\mathrm{M}\ge15~\mathrm{M}_{\mathrm{Jup}}$ prevented us to give an upper limit of the mass.

Lastly, Fig. \ref{fig:col-mag} compares HD\,95086\,b magnitude and color lower-limit with those of other companions, field dwarfs, and tracks from COND and DUSTY evolutionary models (\citealt{baraffe03}, \citealt{chabrier00}). HD\,95086\,b, HR\,8799\,cde, and 2M\,1207\,b appears to be similar in the sense that they all lie at the L -T transition and are less luminous than all other companions but HR\,8799\,b. The Ks - L\,' limit suggests HD\,95086\,b to be at least as cool as HR\,8799\,cde and 2M\,1207\,b. Moreover, with a predicted temperature estimate of $1000\pm 200~$K and log g of $3.85\pm 0.5~$ derived from the L\,' magnitude, HD\,95086\,b would enable further exploration of the impact of reduced surface gravity on the strength of methane bands in the near-infrared.

\subsection{Concluding remarks}

We reported the probable discovery of the exoplanet HD\,95086\,b, which may be the planet with the lowest mass ever imaged around a star.

In summary, our L\,' observations revealed the probable planet in 2012 with a S/N of ten and the likely re-detection of it in 2013 at S/N of three. It is separated from its host-star by $56.4\pm0.7~$AU in projection, has $\mathrm{L\,'}=16.49\pm0.50~$ mag, and an lower-limit for the Ks - L\,' color of $1.2~$mag from the non-detection at Ks. These Ks observations also allowed us to reject the background hypothesis by the most probable contaminants that would have been detected. In addition, we determined the comoving status of the planet with a $3\sigma$ confidence level based on our astrometric measurements in 2012 and 2013. Another dataset of similar quality to the one in 2012 with S/N $\ge 5$ would significantly improve the astrometric precision and thus ascertain the bound status. Finally, we derived a mass of $4\pm1$ to $5\pm1~\mathrm{M}_\mathrm{Jup}$ for HD\,95086\,b using the COND models and an age of $\simeq10$ or $17\pm2~$Myr for the system.

HD\,95086, having a large infrared excess, and its probable planet likely having $q\simeq0.002$, lend support to the assumption that HD\,95086\,b formed within the circumstellar disk like $\beta$\,Pic\,b or HR\,8799\,bcde. Regarding the separation, HD\,95086\,b has a projected physical separation about $56~$AU, which is very similar to the brown-dwarf $\kappa$ And\,b \citep{carson13}, closer to HR\,8799\,b but farther than c and d, and much farther than $\beta$ Pictoris\,b and HR\,8799\,e. Therefore, this giant planet may also be challenging for the classical formation mechanisms, specifically for core-accretion. The timescale to reach the critical core mass of $10~\mathrm{M}_\oplus$ for gas accretion is far longer than the gas dispersal one (from $10^6$ to $10^7~$Myr) and thus prevents the in-situ formation of giant planet beyond very a few tens of AUs \citep{rafikov11}. Particular circumstances for core-accretion \citep{kenyon09} or alternative scenarios like pebble accretion \citep{lambrechts12} may occur instead. Another possible mechanism is gravitational instability. At $56~$AU, the fragmentation of the protostellar disk of HD\,95086 ($1.6~\mathrm{M}_\odot$) can occur \citep{dodson09} but such a low mass planet might not be formed through direct collapse. It may result from subsequent clump fragmentation or from gravitational instability with peculiar disk properties \citep{kratter10}. Finally, there is also the possibility that HD\,95086\,b was formed closer to the star by core-accretion and migrated outwards due to interactions with the disk or to planet-planet scattering \citep{crida09} to its current position. Orbital monitoring showing eccentricity would likely ascertain the presence of an unseen, close-in, and higher mass planet.

Future observations with NaCo and next-generation planet imagers will be important for, first, providing new photometric points at the predicted $H\simeq 18.9~$mag and $K\simeq=18.5~$mag to further explore its atmospheric properties, and second to search for additional close-in planets.




\acknowledgments
The authors would like to thank to the referee for helpful comments and suggestions that improved the quality of this article.
This research has made use of the SIMBAD database operated at CDS, Strasbourg, France; and of the NASA/ IPAC Infrared Science Archive, which is operated by the Jet Propulsion Laboratory, California Institute of Technology, under contract with the National Aeronautics and Space Administration. JR, GC, AML, and PD also thank financial support from the French National
Research Agency (ANR) through project grant ANR10-BLANC0504-01. SD acknowledges partial support from PRIN INAF 2010 ÓPlanetary systems at young agesÓ.



{\it Facilities:}\facility{VLT: Yepun (NaCo)}.

\end{document}